\documentclass[conference]{IEEEtran}
\IEEEoverridecommandlockouts
\usepackage{cite}
\usepackage{amsmath,amssymb,amsfonts}
\usepackage{algorithmic}
\usepackage{graphicx}
\usepackage{textcomp}
\usepackage{xcolor}
\usepackage{url}
\usepackage{todonotes}
\usepackage[inline]{enumitem}
\usepackage[hidelinks]{hyperref}
\usepackage{orcidlink}
\usepackage{subcaption}
\usepackage{pstricks}
\usepackage{pst-node}
\usepackage{pst-text}
\usepackage{pst-grad}
\usepackage{balance}
\usepackage[hang]{footmisc}
\def\BibTeX{{\rm B\kern-.05em{\sc i\kern-.025em b}\kern-.08em
    T\kern-.1667em\lower.7ex\hbox{E}\kern-.125emX}}
\begin{document}

\title{Towards Reproducible Test Annotation for Cyber-Physical Energy Systems using Ontology-driven Dataspaces  %\\

\author{\IEEEauthorblockN{Kai Heussen\IEEEauthorrefmark{1}}
\IEEEauthorblockA{\textit{Dept. of Wind and Energy Systems} \\
\textit{Technical University of Denmark}\\
Roskilde, Denmark\\
0000-0003-3623-1372}
\and
\IEEEauthorblockN{Jawad Kazmi\IEEEauthorrefmark{2}}
\IEEEauthorblockA{\textit{Center for Energy} \\
\textit{AIT Austrian Institute of Technology}\\
 Vienna, Austria \\
0000-0001-9720-6862}
\and
\IEEEauthorblockN{Narges Mehran\IEEEauthorrefmark{3}}
\IEEEauthorblockA{\textit{Salzburg Research GmbH}\\ \textit{\& University of Salzburg}\\
Salzburg, Austria\\
0000-0002-7952-4717}
\and
\IEEEauthorblockN{Artjoms Obushevs\IEEEauthorrefmark{4}}
\IEEEauthorblockA{\textit{Electric Power Systems and Smart Grids} \\
\textit{Zurich University of Applied Sciences}\\
Winterthur, Switzerland \\
0000-0003-3893-4784}
\and
\IEEEauthorblockN{Terence O'Donnell\IEEEauthorrefmark{5}}
\IEEEauthorblockA{\textit{Electrical \& Electronic Engineering} \\
\textit{University College Dublin}\\
Dublin, Ireland \\
0000-0002-2824-7107}
\and
\IEEEauthorblockN{Thomas I. Strasser\IEEEauthorrefmark{2}}
\IEEEauthorblockA{\textit{Center for Energy} \\
\textit{AIT Austrian Institute of Technology}\\
Vienna, Austria \\
0000-0002-6415-766X}
}

% \thanks{K.H. was funded by the Danish Energy Agency, grant no. 134233-51105}
% \thanks{J.K. and T.S. were funded by the EU Horizon Europe research and innovation programme under grant agreement N°101070086.}
% \thanks{N.M. was funded by EXDIGIT project of Land Salzburg under grant no. 20204-WISS/263/6-6022.}
% \thanks{A.O. was funded by Swiss Federal Office of Energy (SFOE) within the Smart Grid International Research Facility Network (SIRFN) cooperative programme on smart grids (ISGAN) under grant agreement SI/502445.}
% }

\thanks{\IEEEauthorrefmark{1} The Danish Energy Agency, grant no. 134233-51105}
\thanks{\IEEEauthorrefmark{2} The EU Horizon Europe research and innovation programme IntNET under grant agreement N°101070086.}
\thanks{\IEEEauthorrefmark{3} The EXDIGIT by Land Salzburg, grant no. 20204-WISS/263/6-6022.}
\thanks{\IEEEauthorrefmark{4} The Swiss Federal Office of Energy under grant agreement SI/502445.}
}

\maketitle

\begin{abstract}
    Reproducibility, traceability, and transparency in testing cyber-physical energy systems are crucial for scientific advancement and cross-laboratory collaboration. Current experimentation and test documentation practices lack formal semantics, making it difficult to reproduce experiments, share data, and apply, for example, the artificial intelligence-driven analysis. A dataspace that relies on structured ontologies aims to address these gaps by providing machine-actionable descriptions. In this work, we outline an ontology-driven approach for reproducibility of cyber-physical energy systems testing and illustrate its applicability through representative cross-laboratory use cases, demonstrating feasibility while identifying remaining semantic and metadata gaps that limit reproducibility. Based on these observations, we propose an open three-viewpoint ontology framework to guide future ontology extensions.
\end{abstract}


\section{Introduction} \label{sec:intro}
Validation of cyber-physical energy systems increasingly relies on complex experimental setups integrating electrical, communication, and control domains. These experiments, ranging from hardware-in-the-loop (HIL) tests to multi-laboratory interoperability trials, are essential for assessing compliance, robustness, and performance under realistic conditions. However, current practices for documenting and sharing such tests remain %largely
informal, relying on narrative descriptions and heterogeneous data formats. This \emph{lack of structured representation} poses significant challenges for reproducibility, transparency, traceability, interoperability, %discovery,
and automation.

%Reproducibility as a Scientific Imperative
Reproducibility is a cornerstone of scientific progress, yet it faces numerous barriers, and the lack of experiment metadata is considered a significant barrier to reproducibility, as is the lack of data itself~\cite{samuel2021understanding}. %metadata is a key element in reprod.
In energy systems research, reproducing an experiment often requires extensive manual interpretation of laboratory notes, configuration files, and procedural documents. Without formalized semantics for test objectives, system configurations, and execution workflows, reproducing results across different laboratories or simulation environments becomes error-prone and resource-intensive.
%A key factor in achieving reproducibility is the creation of the metadata
A key factor in achieving reproducibility is the annotation of rich, machine-readable metadata that systematically captures the context, configuration, provenance, and dependencies of data, software, and experimental workflows, enabling experiments to be reliably comprehended, repeated, and validated across platforms and laboratories. 

%\subsection{Problem statement}
%Need for Structured Annotations in AI Applications
Moreover, the growing use of artificial intelligence (AI) and machine learning in power system analysis amplifies the need for structured metadata. Algorithms that learn from experimental data require a clear context: \textit{what was tested, under which conditions, and according to which criteria}. Ontology-based annotations can provide machine-actionable semantics. %, enabling automated reasoning, feature extraction, and validation of AI models against domain-specific constraints.

%Data Sharing and FAIR Principles
Energy research increasingly adopts \emph{findable, accessible, interoperable, reusable} (FAIR) principles for data management~\cite{wilkinson2016fair}. Ontologies play a pivotal role in achieving interoperability by providing shared vocabularies and formally defined relationships \cite{energy_ontology_survey_2018}. Some of the widely used examples include CIM (Common Information Model) IEC~61970/61968~\cite{iec_cim_2009}, ETSI's SAREF foundation ontology \cite{saref_2017} and its extension for energy SAREF2ENER~\cite{saref4ener_2018}, SEAS (Smart Energy Aware Systems Ontology)~\cite{seas_2015}, Brick (Buildings and Energy Systems) Schema~\cite{brick_2016} and QUDT (Quantities, Units, Dimensions, and Data Types)~\cite{qudt_2014}. A key research gap is the development of structured test descriptions and generic system configuration models enabling datasets to be reproducibly linked to their experimental context. Addressing this gap enhances transparency and facilitates reuse across projects and institutions.
%Missing are structured test descriptions and generic system configuration models, which would enable datasets to be reproducibly linked to their experimental context, thereby improving transparency and reuse across projects and institutions.

%Towards Automated Annotation and Workflow Provenance
Manual annotation of experiments is labour-intensive and prone to omissions and inconsistencies. By introducing ontologies that capture test semantics, system topology, and data context types (e.g., measurement units) combined with provenance models for workflow execution, we aim for semi-automated or fully automated annotation pipelines. This approach aims to support advanced use cases, including compliance checking, reproducibility verification, and cross-domain reasoning.

\subsection{Related Work} \label{sec:related}
%Maybe part of the introduction?
Hacker~\textit{et~al.}~\cite{hacker2024graph} presented a graph-based modelling workflow approach to quantify the impact of cyberattacks in context of distributed energy resources, where grid models are exported to a graph format. %Moreover, the case study is on remotely controllable HEMS systems and the impact on the grid when the central control infrastructure of the HEMS provider gets compromised.
Similarly, Memari and Aljamous~\cite{memari2023model} investigated the feasibility of using PandaPower and Neo4j modelling tools to analyze electrical power systems and to represent grid instances using the CIM ontology classes, emphasizing the generic graph structure in the CIM connectivity data model. Schumilin~\textit{et~al.}~\cite{schumilin2018consistent} identified the Smart Grid domain as the application field for the paradigm of view-based modelling, by using CIM and IEC~61850 as the two central data models of electrical energy systems, noting that CIM only covers the power domain, but that the data model could easily be extended to cover data links as well, thus enabling to accommodate topological information about communication in the same graph. Both \cite{memari2023model} and \cite{schumilin2018consistent} represented the CIM data model as an ontology that explicitly captures grid topology. 

Devanand~\textit{et~al.}~\cite{devanand2020ontopowsys} presented an ontology-based management system capable of formally representing energy management strategies in a wide industrial estate, such as the J-Park simulator. Moreover, cross-domain interactions in a biodiesel plant are studied using their proposed ontology.

Wein~\textit{et~al.}~\cite{wein2024developing} described the ontology development process for research data%within the NFDI4Energy consortium and related it to the five-stage research project life cycle
, emphasizing the central role of the Open Energy Ontology (OEO) \cite{booshehri2021introducing} for their approach. %In particular, the authors introduced and discussed the \emph{energy simulation software ontology}; 
The work highlights the gap regarding reproducibility (of simulated and lab test scenarios) but does not address concrete testing scenarios using metadata.
% they did not elaborate on the testing scenarios and highlight the gap and we push to the gap

Mayer~\textit{et~al.}~\cite{mayer2014ontologies} derives an ontology to document the research infrastructure and processes for conducting e-Science experiments, aligning with the scope of the present paper, but does not apply to an energy system domain. %This paper uses the Wf4Ever Research Object Model, which provides a vocabulary for describing workflow-centric Research Objects by aggregating resources related to scientific workflows.

Horsch~\textit{et~al.}~\cite{horsch2021research} discussed the central role of engineering metadata in enabling sustainable, reusable, and reproducible research data infrastructures. The authors argue that, in addition to storing raw data, research infrastructures must systematically capture rich, machine-readable metadata. % describing models, simulation setups, software, parameters, execution environments, and workflows. However, this work addresses testing and validation pipelines.  %proposed the experiment markup language and elaborated on the metadata development; however, ...

Ford~\textit{et~al.}~\cite{ford2025towards} explored how dataspace technologies can improve data sharing, interoperability, and collaboration in hydrogen research, positioning dataspaces as a key enabler of FAIR data principles, cross-domain collaboration, and reproducible hydrogen research. %while outlining architectural considerations and open challenges for practical adoption. However, this does not annotate metadata using the ontology graph enabled by dataspace. 

%%%%%%\paragraph{Research gap} XXXX
%%%%%%%%%%%%%%%%%%%%%%%%%%%%%%%%%%%%%%%%%%%%%%%%%%%%%%%%%%%%%%%%%%%%%%%
The literature clearly establishes that ontologies provide a foundation for data sharing and that topological ontology models are a key ingredient for grid-related workflows. The need for reproducible simulation and experiment metadata, structured AI-ready annotations, and FAIR-compliant data sharing is motivated and suggested to enable future automation in test documentation and validation workflows. However, the domain does not avail a suitable ontology for this purpose.

The rest of the paper is organized as follows. Section~\ref{sec:background} introduces the background, and Section~\ref{sec:cases} provides an overview of the case studies. The metadata annotation approach supporting reproducibility is detailed in Section~\ref{sec:metadata}, while the overall framework is described in Section~\ref{sec:method}. Finally, Section~\ref{sec:conclusion} concludes the paper.

\section{Background} \label{sec:background}
This section introduces relevant background on testing semantics, ontologies in the energy sector, and the special role of provenance and workflow-oriented annotations, motivating the selection of relevant concepts utilized in the proposed approach.   

\subsection{Testing and Reproducibility of Experiments}
\begin{figure}[t]
    \centering
    \includegraphics[width=0.7\linewidth]{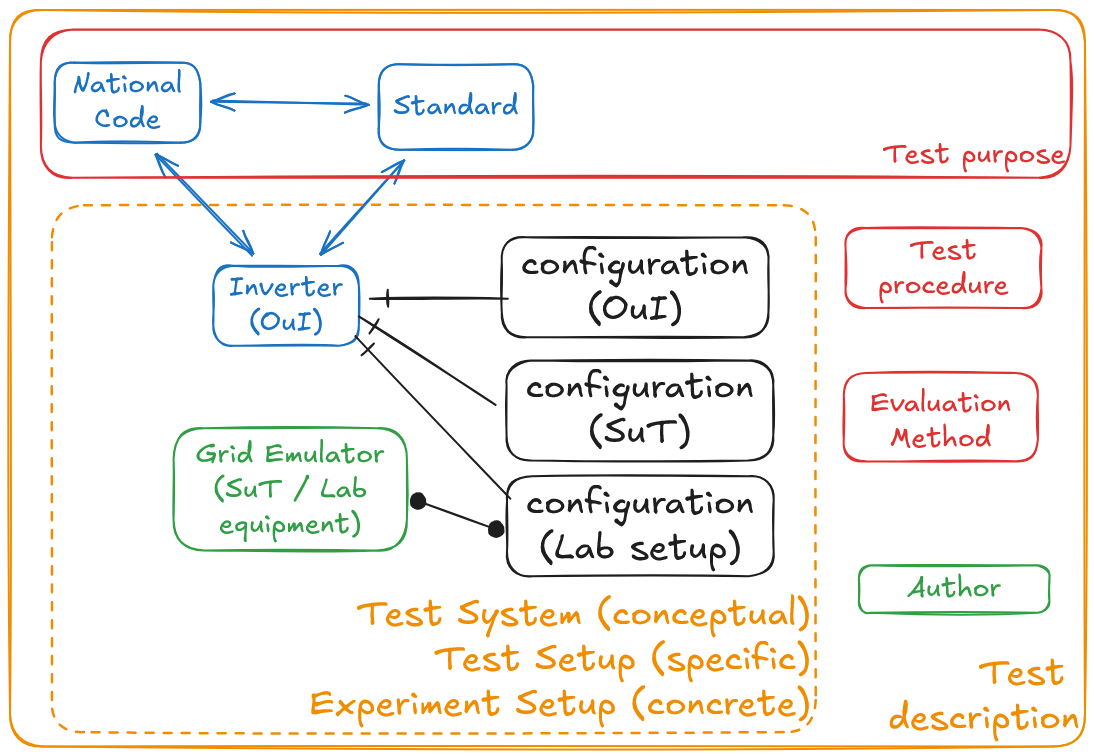}
    \caption{Overview of test description elements, an example of an inverter test (blue/green for example aspects). The orange and red boxes and text refer to elements of HTD, and black identifies different types of system configuration artefacts associated with the test description.}
    \label{fig:test_descr_elements}
\end{figure}

Reproducibility of experiments is a foundational concept across all scientific endeavours. It is therefore not surprising that prior attempts at formalizing experiments exist in the literature across domains. Along this line, the EXPO ontology was proposed in \cite{soldatova2006ontology}, organizing general experiment concepts at a high level, following the principles of the "suggested upper merged ontology" (SUMO), which today is not as widely adopted as, e.g., BFO\footnote{\url{http://ontology.buffalo.edu/bfo/}}. Concrete efforts to achieve end-to-end reproducibility also exist in several domains, but are mainly feasible for computational experiments \cite{mayer2014ontologies, Seaman2023reprodMLops}, where all artifacts are digital. An important framework in testing is the ETSI TTCN-3 standard\footnote{ \url{https://ttcn-3.etsi.org/}}, offering a platform independent framework facilitating automated software testing. For the energy domain, no established testing ontologies exist. Given the complex models and relationships that occur, e.g., in hardware-in-the-loop testing, adoption of ontologies from other domains is difficult. However, the Holistic Test Description (HTD)\footnote{\url{https://github.com/ERIGrid2/holistic-test-description/}} \cite{heussen2017d,heussen2019} offers a documented framework with a range of key concepts, illustrated in Fig.~\ref{fig:test_descr_elements}. The HTD methodology has been adopted \cite{SmILES_D4_3_2019}, further extended \cite{heussen2025uncertainty}, and supplemented with open-source tooling \cite{Schwarz2024toolbox}. While HTD is presented as a set of templates, it spans systematic scope setting, capturing testing intentions, and an essential layering of abstractions: to separate energy systems designs from test designs, and test designs from lab infrastructure considerations. HTD aligns well with established Design of Experiments concepts\cite{van2018design}, which also form the base of above-referenced EXPO. A growing library of fully and partially developed test cases\footnote{\url{https://github.com/ERIGrid2/test-cases}} supplements the method in form of example cases. 

\subsection{Contributions and Approach}
%Objectives of This Work
%\textit{\color{}}
To address the aforementioned challenges and gaps, this empirical work aims to formalize test annotations to facilitate reproducibility. After reviewing related energy test systems, reproduction, and dataspace works, two case studies are presented that focus on reproducing physical and purely digital experiments. The first case study examines metadata annotation. Finally, a modular ontology framework comprising three complementary ontology viewpoints is proposed: (1) \textit{Holistic Test Description Ontology (HTD-O):} %An extension of the Open Energy Ontology (OEO)~\cite{booshehri2021introducing} 
    % On the ontology of holistic test description concepts, 
    formalizing experimental goals, objectives, criteria; %  and their linkage to system configurations and functional views. 
    (2) \textit{System Configuration Model Ontology (SCM-O):} A graph-oriented ontology for representing multi-domain system configuration models;
    (3) \textit{Ontology for Open Provenance Model for Workflow (OPMW) Management:} 
  %  Leveraging the Ontology for Provenance and Workflow Management% (OPMW)
  %  , which extends PROV-O (W3C Provenance Ontology for representing provenance), and P-PLAN (for describing plans, steps, and variables in workflows),
  to capture workflow templates and execution traces for testing activities.%—from lab reservation to data evaluation—and nested sub-sequences (e.g., voltage setpoint profiles).
%\end{itemize}

\subsection{Ontologies in Energy Systems} % Jawad will take care of this section
%\textit{\color{red} work-in-progress (Jawad)} 

Ontologies are formal, machine-interpretable models that define shared concepts, relationships, and constraints for a given domain, enabling a common understanding between humans and machines~\cite{gruber1995ontology, staab2009handbook}. They are typically expressed using standardized semantic web languages such as Resource Description Language (RDF) and Web Ontology Language (OWL), which allow knowledge to be represented as graphs of entities and relations and stored in interoperable formats (e.g., Turtle, XML, N-Triples) that can be queried using languages such as SPARQL (SPARQL Protocol and RDF Query Language))~\cite{hitzler2012owl}. In the energy domain, ontologies have been used to describe power system components, system configurations, measurements, and their units in a consistent, vendor-neutral manner, demonstrating their capabilities to enable interoperability across tools, platforms, and organizations~\cite{iec_cim_2009, saref_2017}. Beyond data integration, ontologies can be utilized to enable automated reasoning, allowing constraints, domain rules, and consistency checks to be formally verified, which is valuable for validating experimental setups, system models, and data in complex cyber-physical energy systems~\cite{qudt_2014, bergenti2018review}. %, analyzed in this work.

\subsection{Provenance and its Applications}
\begin{figure}[t]
    \centering
    \includegraphics[width=0.8\linewidth]{Figures/test_log_provenance.png}
    \caption{Testing process example with provenance concepts: agents (green), activities (red), and entities (black). }
    \label{fig:test_provenance}
\end{figure}
Provenance describes collecting information about the origin, context, and history of data, including how it was produced, by whom, using which processes, and from which inputs \cite{moreau2015prov}. As Fig.~\ref{fig:test_provenance} illustrates, provenance concepts from the W3C PROV-O\footnote{\url{https://www.w3.org/TR/prov-o/}} provide a simple ordering to formally distinguish the relations between  \textit{entities} (physical, digital, conceptual \textit{things}), \textit{activities} (actions, processes, etc.), and \textit{agents} (who carry out an activity) involved in data generation and transformation, and their relationships \cite{prov_dm}. PROV-O is the base ontology that can be used to describe a provenance graph expressed in RDF and OWL. The stored provenance graphs can be stored, queried, and exchanged across laboratories using standard web technologies. In energy and cyber-physical energy systems, provenance plays a crucial role in ensuring traceability, reproducibility, and trust by allowing experimental workflows, simulations, and data processing steps to be transparently documented and validated, thereby supporting data quality assessment, auditing, and compliance verification \cite{missier2013provenance, bergenti2018review}, one of the major focuses of this work.
Open Provenance Model for Workflow (OPMW) extends the W3C PROV model by explicitly capturing the structure and intent of scientific workflows to support reproducibility \cite{garijo2014opmw1}. While PROV records execution-level provenance through {entities}, {activities}, and {agents}, OPMW adds abstractions such as \textit{workflow templates}, \textit{planned processes}, and \textit{variable} roles that describe how an experiment is intended to be performed \cite{opmw_spec}.
%\subsection{Other related topics}

%\subsection{Digital Twin?}

\section{Overview of Case Studies}\label{sec:cases}
% Case studies
We designed two case studies, each targeting a distinct scope of reproducibility. The first case study aimed to reproduce a physical laboratory experiment; the second, a digital-twin-based workflow.  %The test objectives provide relevant context for the documentation and interpretation of results. 
For reproducibility of physical energy systems tests, metadata annotations enable the reconstruction of tests from the original experiments' configurations and logs. In simulation and machine learning experiments, reproducibility primarily concerns identifying data sources consistently and reproducing computational steps. Background data and draft ontologies from the case studies are available \cite{kazmi_2026_18868542}.

\subsection{PV Inverter Test}
\begin{figure}[htbp!]
    \centering
    \subfloat[Conceptual model of test setup with OpenSVP]{\includegraphics[width=.46\columnwidth]{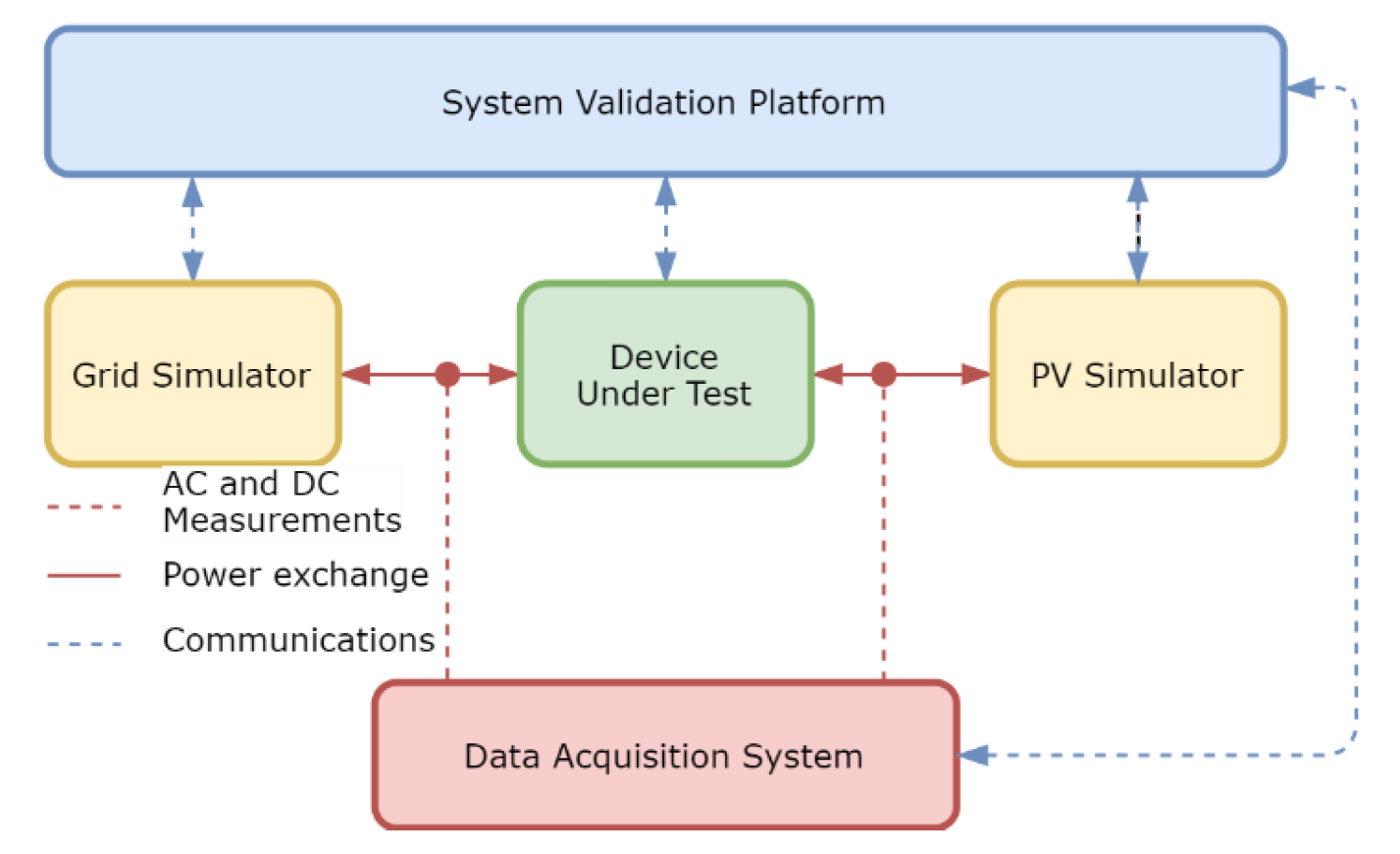}
    \label{fig:inverter}}
    \vspace{.2cm}
    \subfloat[Specific Lab setup in case of UCD experiment]{\includegraphics[width=.46\columnwidth]
    {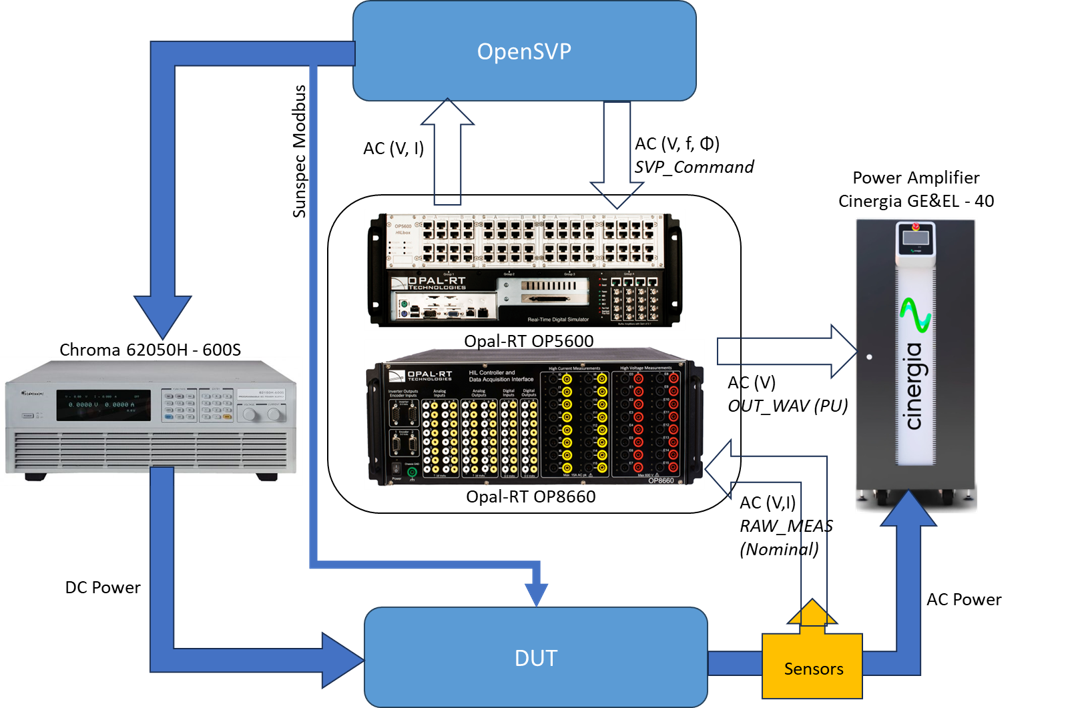}
    \label{fig:inverter_UCD}}
    \caption{Inverter test case study setup.}
\end{figure}

%  Inverter test
The photovoltaic (PV) inverter test case study, depicted in Figures~\ref{fig:inverter}--\ref{fig:inverter_UCD}, is part of an experiment setup at the UCD institute in Ireland, consisting of the following devices:
\begin{enumerate*}[label=\textit{\roman*)}]
    \item One 3-phase grid simulator.
    \item One DC amplifier.
    \item One RT computer/data logger.
    \item Voltage and Current Sensors.
    \item One PV inverter as the System under Test (SuT).
\end{enumerate*}
A grid simulator provides stable alternating circuit (AC) conditions for the photovoltaic (PV) inverter and absorbs the inverter's output power. A direct current (DC) amplifier emulates the PV array on the DC side controlled by a computer that applies selectable I-V characteristics based on module type, irradiance, temperature, and shading. The same computer monitors real-time DC voltage/current and visualizes the operating point against the theoretical I-V curve. EN~50549 grid support function tests are defined by the user for selected parameters and measured using sensors from the AC sides with a 1s instantaneous window.

\subsection{Digital-twin-based Workflow}
\begin{figure}[htbp!]
    \centering
    \includegraphics[width=\columnwidth]
    {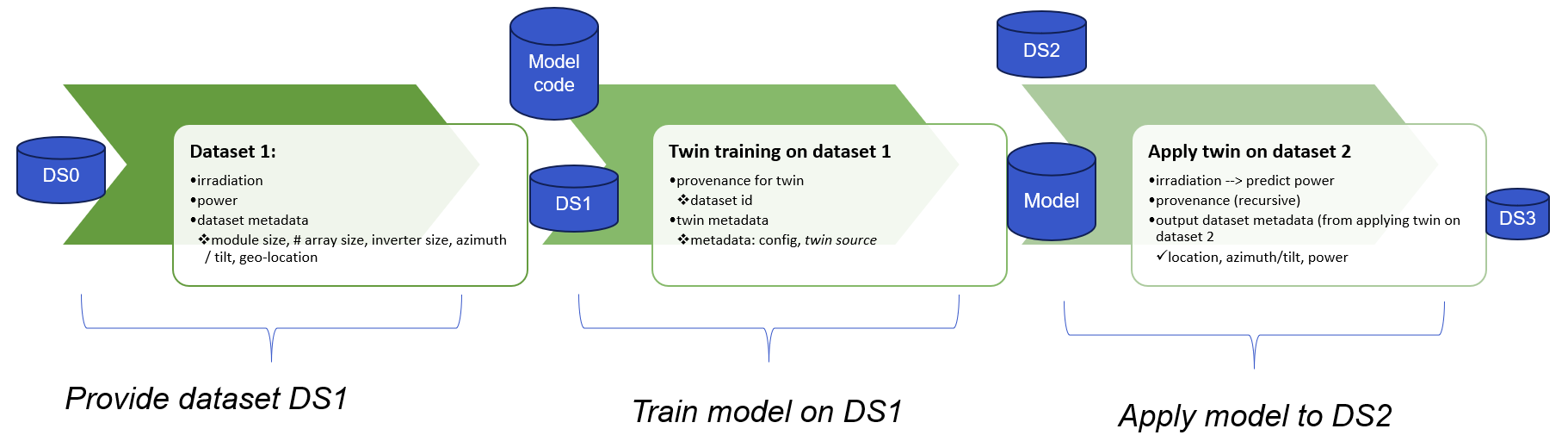}
    \caption{Digital-twin-based workflow.}
    \label{fig:DT}
\end{figure}

%  Digital-twin-based Workflow 
As shown in Fig.~\ref{fig:DT}, this use case demonstrates a digital twin–based workflow for PV power estimation with full provenance tracking. A well-characterized reference dataset (DS1), containing irradiance, measured power, and detailed system metadata, is used to train a PV digital twin while recording the dataset, model configuration, and code provenance. The trained twin is then applied to a second dataset (DS2), where only irradiation and basic site parameters are available, to predict PV power and generate an enriched output dataset (DS3). 
As a digital process with identified entities, its steps should be well reproducible. The replicability of this process is a case for provenance as it requires %that provenance information is propagated recursively,
linking predictions back to the original data and model. %, enabling scalable, reusable, and traceable PV power estimation across locations without requiring additional physical measurements.

%\subsection{Further Cases}
%  Further cases ... multi-domain simulation, e.g PHIL lab experiments (HTD layers), 
%Further cases include multi-domain simulation scenarios, such as power hardware-in-the-loop (PHIL) laboratory experiments, where cyber, control, and physical power-system components interact across multiple abstraction layers. In these settings, heterogeneous models (e.g., power electronics, communication networks, control algorithms, and physical grid components) are coupled in real-time, often spanning hardware, software, and simulation of HTD layers. Such experiments introduce additional challenges for reproducibility and digitalization, as they require consistent metadata for hardware configurations, real-time interfaces, timing constraints, and cross-domain dependencies, making them a demanding use case for metadata-enabled, ontology-driven experimentation.

\section{Metadata Annotation and Reproducibility} \label{sec:metadata}
Regarding the PV inverter test case study, this section describes how we applied an existing dataspace ontology to annotate metadata from a facilitated test of an OpenSVP software suite\footnote{\url{https://github.com/sunspec/svp}}, and what limitations we observed for reproducibility from an energy system test conducted with the open-source ~\cite{odonnell2024automated}. Thereafter, we discuss the test reproducibility by the ZHAW institute in Switzerland, based on the original test conducted by the UCD.

\subsection{Test Metadata Annotation from OpenSVP}
We used a basic dataset available from the test conducted by the software components of OpenSVP~\cite{johnson2018international}: test design/specification level (i.e., test case, suite), executable software artifacts (i.e., script, library, and processing software), and runtime and execution artifacts (i.e., result, log entry). 
The data model described was based on the metrics measured/configured and the data collected.
Thereafter, we modelled our ontology and stored it in the default outline of the \textit{turtle document format} (\texttt{*.ttl}), allowing us to represent an RDF graph in a compact textual form. 

\begin{figure}[htbp!]
    \centering
    \subfloat[Metadata modelling and annotation process.]{\includegraphics[width=.48\columnwidth]
    {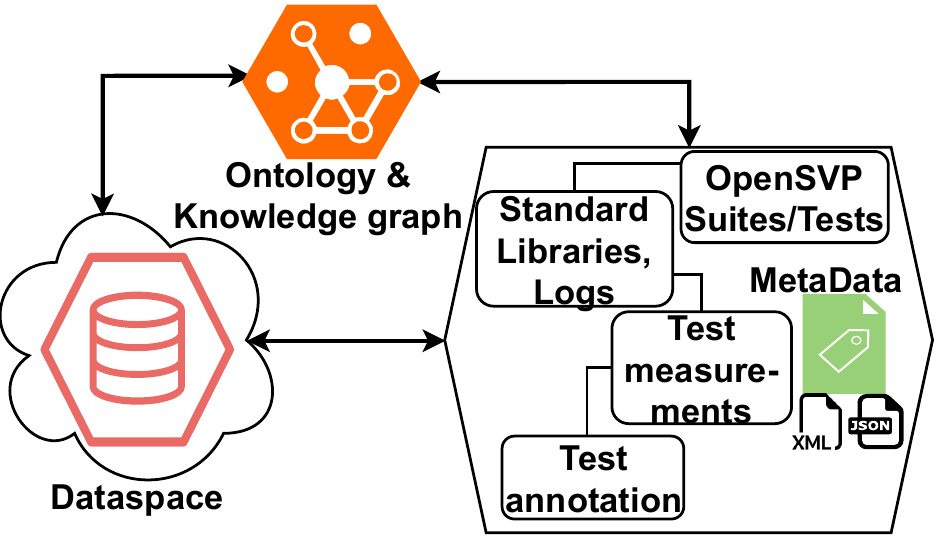}
    \label{fig:metamodel}}\vspace*{.2cm}
    \subfloat[Data model for OpenSVP case study.]{\includegraphics[width=.48\columnwidth]
    {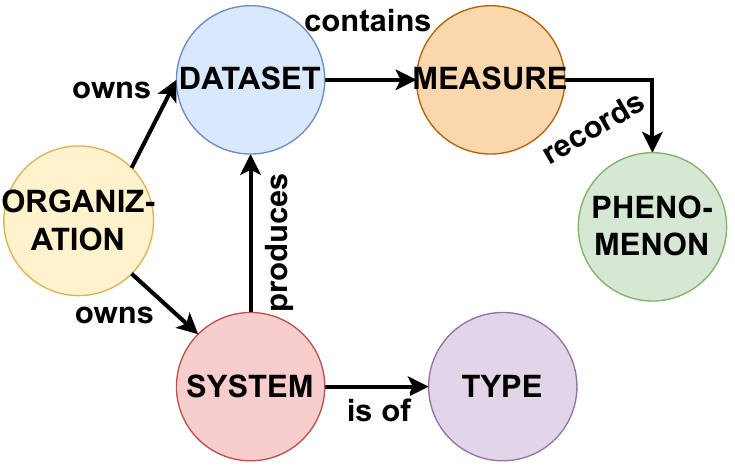}
    \label{fig:Ontology-graph}}
    \caption{Overview of metadata modelling and ontology graph.}
\end{figure}
As shown in Fig.~\ref{fig:metamodel}, the objective was to extract a metadata file (e.g., \textit{XML}, \textit{JSON}) associated with the ontology components and the dataspace of our consortium. In detail, the dataset received from the OpenSVP test comprises measurements of AC voltage, current, active power, and reactive power. First, we investigated logs provided by UCD, which are ordered sequences of test operations associated with a test suite and contain multiple test cases and, optionally, nested test suites. We then extracted the custom entities and required information from the measurements and phenomena defined in the OpenSVP dataset to ensure test reproducibility and to create an ontology integrated into the knowledge graph in the dataspace. 
\begin{figure}[htbp!]
\centering
    \includegraphics[width=\columnwidth]
    {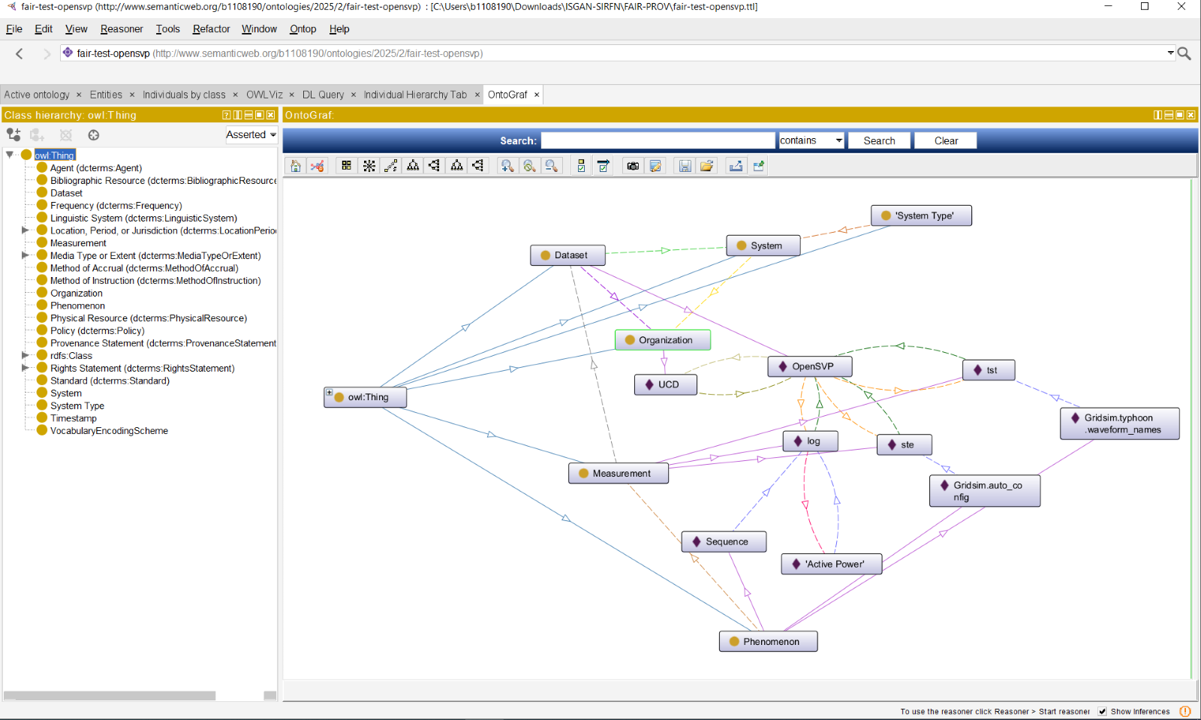}
    \caption{Ontology graph of test annotation in the open-source Protégé tool.}
    \label{fig:Onto-in-Protege}
\end{figure}
Fig.~\ref{fig:Ontology-graph} shows that, firstly, UCD, which owns an energy test system and an OpenSVP dataset, provides it for the case of reproducibility by ZHAW. Moreover, each system is of a specific type (e.g., PV system, switchboard, AC power grid). This dataset contains multiple log files that store measurements of the test setup. Last but not least, each measurement records a specific phenomenon, such as active power or voltage. 

Fig.~\ref{fig:Onto-in-Protege} shows an extended ontology modelled in the open-source tool Prot{\'e}g{\'e}~\cite{10.1145/2757001.2757003}, in accordance with the customized entities defined earlier. To elaborate, there are multiple terms related to OpenSVP configurations, such as the OpenSVP suite configuration file (STE) and the test configuration file (TST).

%Thereafter, we generated the ontology graph using custom entities from OpenSVP and extracted metadata from the measurements and phenomena defined therein.
% \begin{enumerate}
%     \item Classes: Organization, Dataset, Measurement, and Phenomenon;
%     \item Object Properties: ownsDataset, containsMeasurements, recordsPhenomenon;
%     \item Individuals: UCD, OpenSVP, suite (STE), test (TST), log, active power, power steps;
%     \item Data Properties: hasSTEFile, hasTSTFile, hasLogFile, hasMeasurementUnit.
% \end{enumerate}

\subsection{Reproducibility of EN~50549 OpenSVP-based Test}

\begin{figure}[htbp!]
    \centering
    \subfloat[Normal Operation Range]{\includegraphics[width=.95\columnwidth]
    {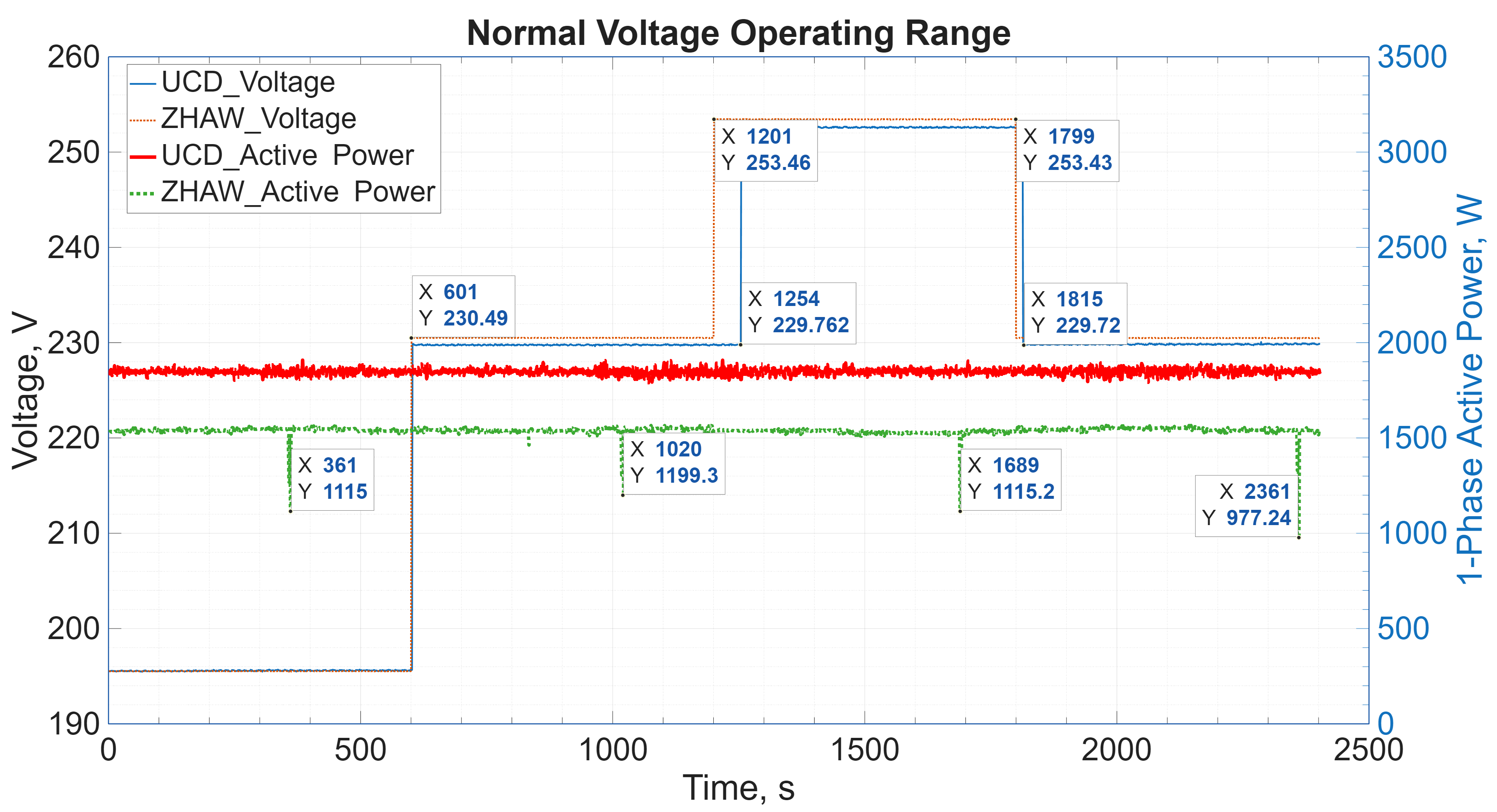}
    \label{fig:NOR}}\vspace*{.2cm}
    \subfloat[Active Power Reduction]{\includegraphics[width=.95\columnwidth]
    {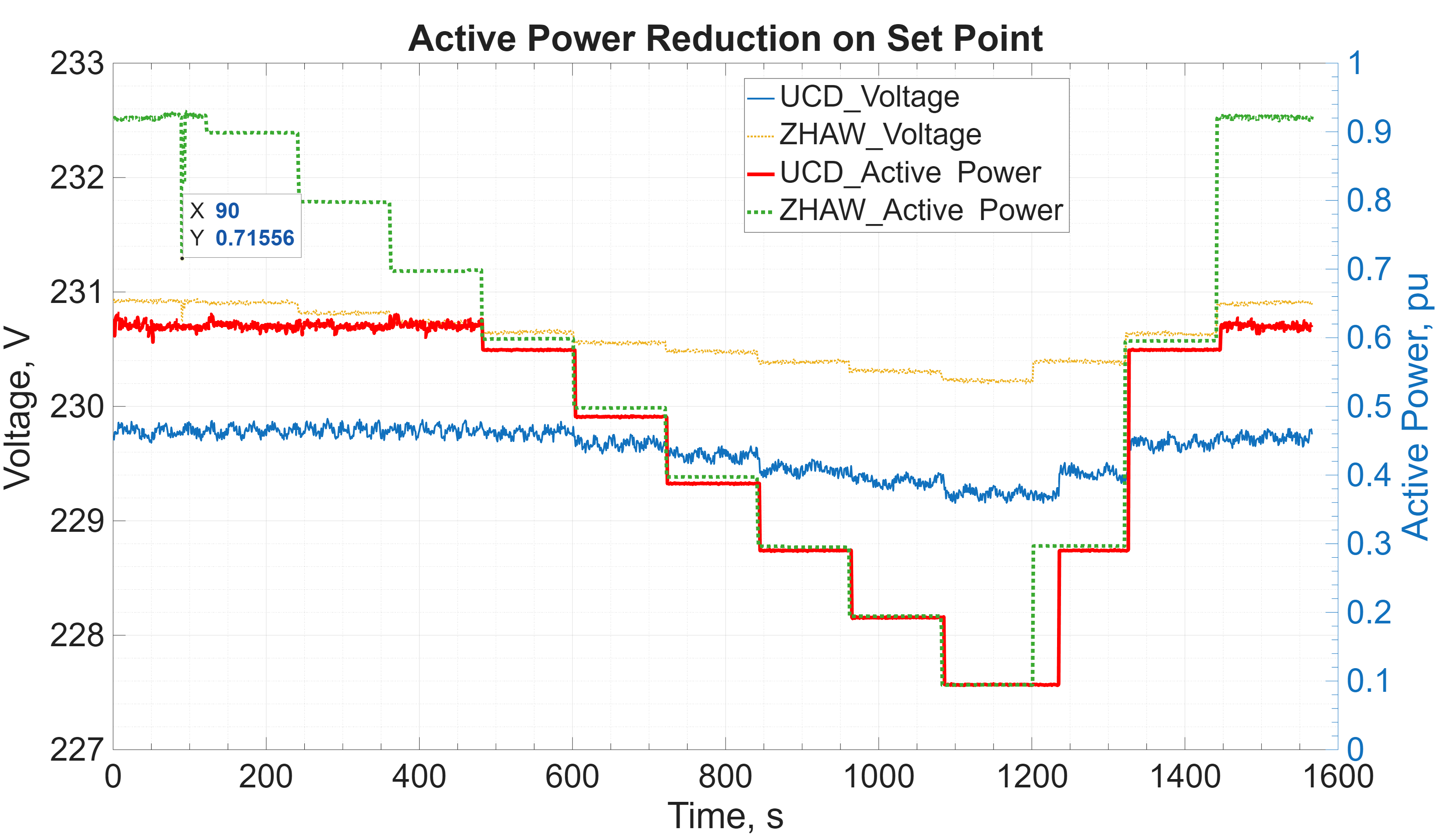}
    \label{fig:ARP}}
    \caption{Reproduction of EN 50549-10:2022 tests.}
\end{figure}

The steps below reproduce the original test sequence, settings, and setpoints from the tests conducted at UCD. OpenSVP testing activities include EN~50549-10:2022 tests for the conformity assessment of generating units, focusing on three-phase PV inverters at ZHAW, Switzerland (the original tests at UCD were conducted on a single-phase infrastructure), including interactions with national interface-protection settings. The test comprises inverter configurations, test procedures, and grid-emulator programming, while also highlighting gaps such as missing detailed ramping procedures and limited formal documentation of the test setups. Measurements primarily capture voltage and power profiles over time, with only partial recording of currents and reactive power, and they lack explicit logging of critical events (e.g., the inverter breaker state). Test evaluation was performed against the applicable standards using pass/fail criteria (e.g., ride-through versus disconnection). Fig.~\ref{fig:NOR} shows reproduction case for the continuous voltage operating range: the initial voltage setpoint is at the lower limit, 85\% of $U_n$, and after 600~s it is increased to 100\% and then to the upper limit, 110\%. Fig. \ref{fig:ARP} shows the corresponding active power reduction sequence: the UCD single-phase inverter operates at 62\% and the ZHAW three-phase inverter at 92\%, with set-point changes applied in 120~s steps. The active power set point is decreased from 0.9~pu to 0.1~pu of nominal power in steps of 0.1~pu. Once 0.1~pu is reached, the set point is increased to 0.3~pu, then to 0.6~pu and finally to 1.0~pu, applying the 1~min average active power evaluation procedure for each level.

Both inverters configured with national settings passed the tests; however, specific discrepancies were identified, highlighting the need for consistent, complete setup documentation. Improvements are needed in impedance characterization, metadata capture, and more systematic recording to support reproducibility and automated post-processing.

\section{Proposed Framework for Ontology-based Annotation for Experiment Reproducibility} \label{sec:method}
% (viewpoints and overall idea)
% Methodology (viewpoints and overall idea) -> 
The conceptual framework for reproducible testing data entails both a procedural and a semantic understanding of the testing context, comprising a test case, a laboratory configuration, and the integration of the device or function being tested into the laboratory testbed. To facilitate machine-readable and consistent annotation of these complex relations, we propose a combination of ontology-based annotations to address and disambiguate the various contexts.   

To formally capture the dimensions of a testing process listed above, we propose several descriptive viewpoints, each supported by a suitable ontology. The following subsection outlines these descriptive viewpoints together with their respective ontologies, after which we illustrate the conceptual reproduction workflow that would be enabled by the proposed annotations. We assume an ontology-enabled dataspace that enables semantic queries across research infrastructures, facilitated by the proposed ontologies.  

\subsection{Descriptive viewpoints for Test Reproducibility}
% The modeling of systems is per ...
%  of the testing workflows and the

% HTD-O, SCM-O, %OPWM
% OPMW?

The modeling of the framework is proposed using a three-viewpoint approach. This way, it captures the conceptual and methodological description of a test using \mbox{HTD-O}, while \mbox{SCM-O} formalizes the concrete system configuration to which the test is applied, and \mbox{OPMW} represents the workflow structure that operationalizes the test through processes, data dependencies, and parameter bindings. A more detailed description of each layer follows below.

%  Subsection on 3-layer concept
\paragraph{Holistic Test Description Ontology (HTD-O)}
%The open energy ontology offers a context and platform for hosting energy systems-related ontologies. 
%The holistic test description REF HERE has been developed for the broad context of testing in energy systems, including cross-domain testing, we consider that the OEO would suit as a   
Based on the well-documented holistic test description methodology, we propose an ontology view that would offer the HTD terms and establish the collections of concepts associated with the HTD description layers. The HTD is internally consistent and offers good coverage of the testing domain, HTD-O will start with a smaller set of key concepts, simpler for annotations. HTD facilitates the association with relevant system configuration definitions, as illustrations or formal definitions using SCM-O.
The HTD ontology should align with the high-level concepts from EXPO\footnote{\url{https://expo.sourceforge.net/}}~\cite{soldatova2006ontology}, without adopting its upper ontology.

\paragraph{System Configuration Model Ontology (SCM-O)} %Separate from the HTD concepts, but also introduced in \cite{heussen2017d}, is a specific notation of system configurations. 
Although several ontologies exist for modelling systems, each has been developed in its respective domain. However, in \cite{heussen2017d}, a simple topology-oriented high-level description of the system configurations, based on systems and domain-specific connections, was proposed.

\begin{figure}[htbp!]
    \centering
    \includegraphics[width=1.0\linewidth]{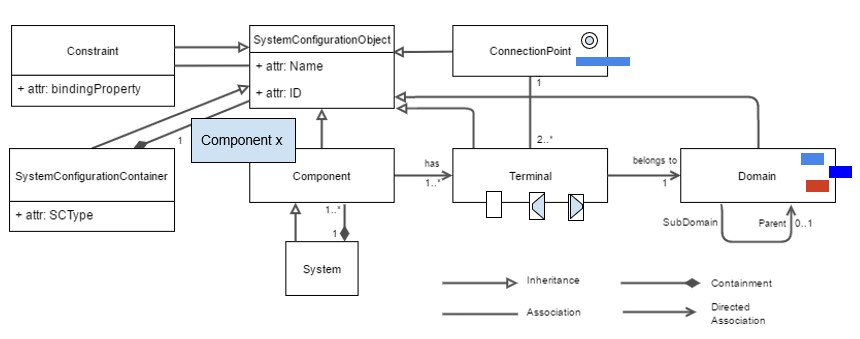}
    \caption{Outline of System configuration model ontology (SCM-O) and diagrams (SCM-D). Diagram reproduced from \cite{heussen2017d}.}
    \label{fig:SCM}
\end{figure}

The data model, shown in Fig.~\ref{fig:SCM}, defines the essential structural elements for topologically modelling system configurations. The data model provides a generic multi-domain representation that supports algorithmic reasoning and is expected to align well with existing standard configuration representations such as CIM, IEC61850-6 (SCL), SysML,  SEAS and SAREF4SYST. Alignment with CIM, SCL and SysML was discussed in \cite{heussen2017d}, but has not yet been assessed with respect to SEAS and SAREF4SYST.

Here, we the formulation of SCM-O as a lightweight ontology to model relevant system configuration information for the testing context, support it with digram tooling and mappings from the aforementioned standards. Complementarity with HTD-O would be given by-design.

%FIGURE: D5.2 data model + maybe example of layering.

%Fig.~\ref{fig:viewpoints} ...

\paragraph{OPMW}
Neither HTD-O, nor SCM-O offer a procedural view. The practical workflow illustrated in Fig.~\ref{fig:test_provenance} shows, on the one hand, that the basic PROV-O concepts apply well to describing workflows. Since typical testing and laboratory workflows follow a reference and are embedded (e.g., a test procedure is embedded in the illustrated testing workflow), it seems appropriate to adopt stronger workflow semantics. OPMW offers these capabilities by separating workflow specification from execution. OPMW enables the representation of both \textit{prospective} and \textit{retrospective} provenance, facilitating experiment validation, comparison, and reuse.

%  Ontology-enabled Datasapce Workflow for reproduction (Jawad)
%\textit{\color{red} work-in-progress (Jawad)} 
\subsection{Workflow for Reproduction using an Ontology-enabled Dataspace}
As explained in detail in Section~\ref{sec:method}, the proposed framework, supported by a dataspace infrastructure, provides the semantic glue among high-level test goals, system configuration, and executable test procedures. 
%Together, these ontological viewpoints enable the systematic linkage of test definitions, system instances, parameter settings, and resulting datasets, supporting validation, comparison, and reliable reproduction of cyber-physical energy system tests. %\cite{garijo2014opmw2, garijo2014towards,prov_o}.
Fig.~\ref{fig:metamodel} %\textcolor{red}{All: please check if you agree with this reference}
provides a conceptual overview of the annotation and discovery workflow in a dataspace context. First, system configuration, test objectives, and execution procedure are jointly described using ontology-based RDF/OWL representations, ensuring machine-interpretable definitions of inputs, outputs, constraints, parameters, and process dependencies across all viewpoints. %This representation captures the conceptual test definition, the system-under-test instantiation, and the specification of the planned test structure. 
Second, all semantic artifacts and related datasets are \textit{published} to the dataspace infrastructure, where they are stored in a \textit{triplestore} and made discoverable through standardized interfaces. Finally, dataspace services are used to query, analyze, and share test definitions, configurations, execution traces, and results. %, enabling stakeholders to validate hypotheses, compare outcomes, and reliably reproduce the test under similar or modified setups.
%\textcolor{red}{Jawad - we need somthing more here.}

%\begin{figure}[htbp]
%    \centering
%    \includegraphics[width=.9\columnwidth]{Figures/FAIRTEST_SCM-D_layers.jpg}
%    \caption{System configuration layers expressed in SCM-O.}
%    \label{fig:layers}
%\end{figure}

\section{Conclusion and Outlook}\label{sec:conclusion}
% Conclusion
To reproduce energy test system experiments conducted across two different laboratories, this paper studies metadata for test annotation and proposes a set of ontologies to formalize the demonstrated annotations. 
To enable the validation of the test annotations for reproducibility, we designed two case studies on reproducing physical and purely digital experiments. %For the inverter test case study, we presented the metadata annotation and reproduction steps and metadata gaps identified using our dataspace-based annotation approach. 
Finally, we propose a modular ontology framework comprising three interrelated ontology viewpoints. 
%% Ontology-enabled dataspace proof of concept (Jawad)
%%%%%%%%%%%%%%%%Ontology-enabled dataspace  - Datspace
%% Federated reproducibility arrangements using OeDS, for testing and validation
% Outlook
%% Tool integration and automation - especially for configuration-driven simulation and testing pipelines (Design of Experiments, PSAL - automated cyber-physical testing, ..)
%% Annotation tools (potential for workflow-oriented automated annotation, AI tools) --> enable workflows
%% Templates, (potentially a library of templates ) for default workflows and experiment types ... 
Future work will focus on formalizing and exemplifying the proposed draft ontologies in an open repository, a deeper tool integration and automation, particularly for configuration-driven simulation and testing pipelines such as Design of Experiments (DoE) and automated cyber-physical testing using approaches like the Power System Automation Language (PSAL). Workflow-oriented and AI-assisted annotation support tools are expected to be key to the practical acceptance of the test annotation and ontology utilization. %In addition, reusable templates and a growing library of default workflows and experiment types will be developed to support consistency and rapid setup. 
The approach will be further strengthened through an ontology-enabled dataspace proof of concept and federated reproducibility arrangements, thereby enabling distributed testing, validation, and reproducible experimentation across organizational boundaries.

% \begin{table}[htbp]
% \caption{Table Type Styles}
% \begin{center}
% \begin{tabular}{|c|c|c|c|}
% \hline
% \textbf{Table}&\multicolumn{3}{|c|}{\textbf{Table Column Head}} \\
% \cline{2-4} 
% \textbf{Head} & \textbf{\textit{Table column subhead}}& \textbf{\textit{Subhead}}& \textbf{\textit{Subhead}} \\
% \hline
% copy& More table copy$^{\mathrm{a}}$& &  \\
% \hline
% \multicolumn{4}{l}{$^{\mathrm{a}}$Sample of a Table footnote.}
% \end{tabular}
% \label{tab1}
% \end{center}
% \end{table}

% \begin{figure}[htbp]
% \centerline{\includegraphics{Archive/fig1.png}}
% \caption{Example of a figure caption.}
% \label{fig}
% \end{figure}

% Figure Labels: Use 8 point Times New Roman for Figure labels. Use words 
% rather than symbols or abbreviations when writing Figure axis labels to 
% avoid confusing the reader. As an example, write the quantity 
% ``Magnetization'', or ``Magnetization, M'', not just ``M''. If including 
% units in the label, present them within parentheses. Do not label axes only 
% with units. In the example, write ``Magnetization (A/m)'' or ``Magnetization 
% \{A[m(1)]\}'', not just ``A/m''. Do not label axes with a ratio of 
% quantities and units. For example, write ``Temperature (K)'', not 
% ``Temperature/K''.

\section*{Acknowledgment}

The authors acknowledge the fruitful collaboration and facilitation provided by the Smart Grid International Research Facility Network (SIRFN) within the Implementing Agreement for a Cooperative Programme on Smart Grids (ISGAN), which operates as both an IEA Technology Collaboration Programme and a Clean Energy Ministerial initiative.

%This work acknowledges the Swiss Federal Office of Energy (SFOE) within the Smart Grid International Research Facility Network (SIRFN) cooperative programme on smart grids (ISGAN) under grant agreement No. SI/502445 and the Excellence in Digital Sciences and Interdisciplinary Technologies (EXDIGIT) project funded by Land Salzburg under grant number 20204-WISS/263/6-6022. %\textit{\color{red} int:net project? (Jawad, Thomas)} 

% The preferred spelling of the word ``acknowledgment'' in America is without 
% an ``e'' after the ``g''. Avoid the stilted expression ``one of us (R. B. 
% G.) thanks $\ldots$''. Instead, try ``R. B. G. thanks$\ldots$''. Put sponsor 
% acknowledgments in the unnumbered footnote on the first page.

\bibliographystyle{IEEEtran}
\balance
\bibliography{ref}

%%%%%%%%%%%%%%%%%%%%%%%%%%%%%%%%%%%%%%%%%%%%%%%%%%%%%%%%%%%%%%%%%%%%%%%
% Introduction + Motivation  [Ontology-enabled Dataspace OeDS], interoperability [at knowledge level], boosting discoverability and data  quality
%+ Problem Statement + our contribution (All) + related work
% Background on testing + ontologies (Jawad) + other related topoics
% Methodology (viewpoints and overall idea) -> 
%  Subsection on 3-layer concept
%  Individual Subsection on each layer
%  Ontology-enabled Datasapce Workflow for reproduction (Jawad)
% Case studies
%  Inverter test
%  Digital-twin-based Workflow 
%  Further cases ... multi-domain simulation, e.g PHIL lab experiments (HTD layers), 
% Outlook
%% Tool integration and automation - especially for configuration-driven simulation and testing pipelines (Design of Experiments, PSAL - automated cyber-physical testing, ..)
%% Annotation tools (potential for workflow-oriented automated annotation, AI tools) --> enable workflows
%% Templates, (potentially a library of templates ) for default workflows and experiment types ... 
%% Ontology-enabled dataspace proof of concept (Jawad)
%% Federated reproducibility arrangements using OeDS, for testing and validation
%%%%%%%%%%%%%%%%%%%%%%%%%%%%%%%%%%%%%%%%%%%%%%%%%%%%%%%%%%%%%%%%%%%%%%%

\end{document}